\documentclass[prl,superscriptaddress,showpacs,preprintnumbers,amsmath,amssymb,twocolumn,floats]{revtex4}
\usepackage{txfonts}
\usepackage{amssymb}
\usepackage{graphicx}
\usepackage{sidecap}
\usepackage{CJK}
\usepackage{txfonts}

\begin{document}

\title{Angle-resolved Photoemission Spectroscopy Study on the Surface States of  the Correlated Topological Insulator YbB$_6$}

\author{M. Xia \footnote{These authors contribute equally to this work.}}

\author{J. Jiang \footnote{These authors contribute equally to this work.}}

\author{Z. R. Ye }

\affiliation{State Key Laboratory of Surface Physics, Department of Physics, and Advanced Materials Laboratory, Fudan University, Shanghai 200433, China}

\author {Y. H. Wang}

\affiliation{Department of Physics and Applied Physics, Stanford University, Stanford, California, 94305, USA}

\author{Y. Zhang}

\affiliation{SLAC National Accelerator Laboratory, Stanford Institute for Materials and Energy Sciences,
2575 Sand Hill Road, Menlo Park, California 94025, USA}

\author{S. D. Chen}

\affiliation{Department of Physics and Applied Physics, Stanford University, Stanford, California, 94305, USA}

 \author{X. H. Niu }

\author{D. F. Xu  }

\affiliation{State Key Laboratory of Surface Physics, Department of Physics, and Advanced Materials Laboratory, Fudan University, Shanghai 200433, China}

\author{F. Chen}
\author{X. H. Chen}
\affiliation{Department of Physics and Hefei National Laboratory for Physical Science at Microscale, University of Science and Technology of China, Hefei 230026, China.}

\author{B. P. Xie }

\author{T. Zhang }

\email{tzhang18@fudan.edu.cn}

\author{D. L. Feng }

\email{dlfeng@fudan.edu.cn}

\affiliation{State Key Laboratory of Surface Physics, Department of Physics, and Advanced Materials Laboratory, Fudan University, Shanghai 200433, China}


\begin{abstract}

We report the electronic structure of YbB$_6$, a recently predicted moderately correlated topological insulator, measured by angle-resolved photoemission spectroscopy. We directly observed linearly dispersive bands around the time-reversal invariant momenta $\overline{\Gamma}$ and $\overline{X}$ with negligible $k_z$ dependence, consistent with odd number of surface states crossing the Fermi level in a Z$_2$ topological insulator. Circular dichroism photoemission spectra suggest that these in-gap states possess chirality of orbital angular momentum, which is related to the chiral spin texture, further indicative of their topological nature. The observed insulating gap of YbB$_6$ is about 100~meV, larger than that reported by theoretical calculations. Our results present strong evidence that YbB$_6$ is a correlated topological insulator and provide a foundation for further studies of this promising material.

\end{abstract}

\maketitle

Topological insulator (TI) is a new class of matter with topologically protected surface states that possess unique electronic and spin properties. Recently, how the topological order interplays with the electronic correlation has attracted a lot of theoretical considerations. As materials with significant correlations, the rare-earth borides are interesting with a variety of correlated phenomena including mixed valence, heavy fermion and superconductivity \cite{ReB6,valence,fermion,metal,semi,super,spin,TEM}. Especially, samarium hexaboride (SmB$_6$) has recently been predicted to be a topological Kondo insulator \cite{Sm1, Sm2}, which fueled intense effort in search for topological orders in correlated systems. Numerous experiments have been performed to identify the topological surface states in SmB$_6$\cite{JJSmB6, HasanSmB6, PRXSmB6}. However, the surface states can only exist in low temperature and the coexistence of the bulk bands and surface states in SmB$_6$ limits its future applications.

\begin{figure*}[t]
\includegraphics[width=18cm]{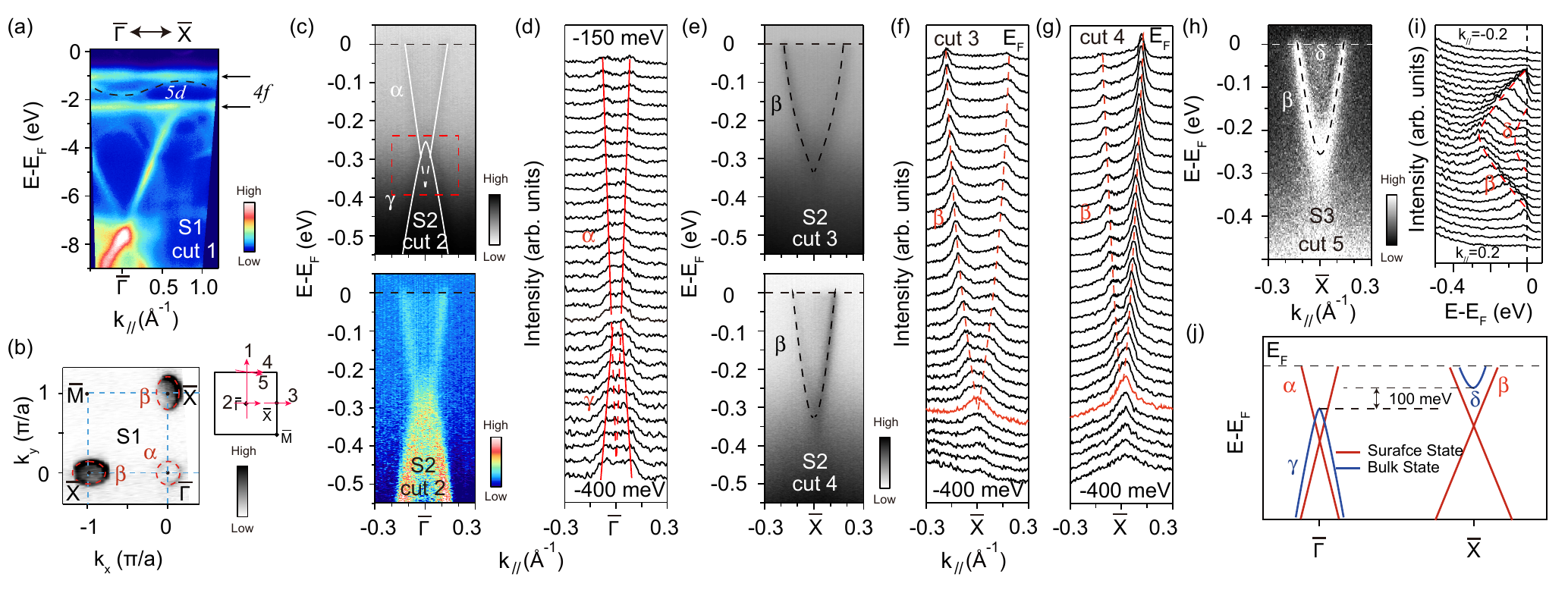}
\caption{(Color online) (a) Photoemission intensity plot showing the valence band structure of YbB$_6$. The flat 4$f$ bands are marked by the arrows and the 5$d$ band is indicated by the dashed line. Data were taken along cut 1 of S1 at 6~K with 33 eV photons, at SSRL. (b) Left: a photoemission intensity map at $E_F$ for YbB$_6$ taken with 31~eV photons at 6~K in SSRL.  The intensity was integrated over a window of [$E_F$ - 10~meV, $E_F$ + 10 ~meV].
Right: the momentum cuts along which data were taken are marked in the projected two-dimensional BZ . (c) Top: the photoemission intensity plot along cut 2 taken on S2 with 91~eV photons at 18~K in SLS. Bottom: the same data after  subtracting  a momentum-independent background. The spectra far away from the dispersive region show no momentum dependence and thus taken as the background. The white lines indicate the $\alpha$ and $\gamma$ bands, the dashed line is extrapolated from the linear dispersion. The red dashed line indicates the overlap of the band bottom of $\alpha$ and band top of $\beta$. (d) The corresponding MDCs along cut 2 within [$E_F$ - 400~meV, $E_F$ - 150 ~meV]. The red  line is the dispersion tracked from the MDC peaks and the red dashed line is extrapolated from the linear dispersion. (e) The photoemission intensity plot on S2 along cut 3 and cut 4 measured with 70~eV photons at 18~K in SLS. (f), (g) The corresponding MDCs along cut 3 and cut 4 in panel (e), respectively. The red dashed lines are the dispersions of the $\beta$ bands. The highlighted MDC in red indicates the possible energy position of the Dirac points. (h), (i) The photoemission intensity plot on S3 along cut 5 and its corresponding energy distribution curves (EDCs), showing an additional $\delta$ band. Data were taken with 47~eV photons at 20~K in KEK. (j) A schematic of all the bands observed along $\overline{\Gamma}$-$\overline{X}$ direction. The detailed description is in the text.
} \label{cut}
\end{figure*}

YbB$_6$ is a related rare-earth hexaborides which shares the same CsCl type crystal structure with SmB$_6$ \cite{CsCltype}, and is predicted to be a correlated TI with a bulk insulating gap of about 31~meV  \cite{Ybcalc}.  As it is not a Kondo insulator, and the gap is comparable to ambient temperature,  it would be  more suitable for applications based on topological surface states  than SmB$_6$. However, despite the previous  successes of density functional theory in predicting TIs  \cite{HgTe1, HgTe2,BiSe1, BiSe2, BiSe3}, the strong correlations in rare-earth compounds pose challenges to an accurate calculation. For instance, the resistivity of  YbB$_6$  exhibits a bulk metallic behavior \cite{Res}, inconsistent with the prediction.
Therefore, it is crucial to experimentally determine the electronic structure and topological nature of YbB$_6$. Angle-resolved photoemission spectroscopy (ARPES) is a powerful tool to directly measure the surface and bulk band dispersions and determine the value of topological invariant of three-dimensional TIs \cite{BiSe2,doping1,doping2,HasanSmB6}. Furthermore, circular dichroism (CD) of the ARPES data have been shown to be related to the spin textures of the surface states \cite{JJSmB6,BiTeCD, YHWang, CD_zhu, Louie, Changyoung, YHWang_2}.

In this Letter, we report our ARPES measurements on YbB$_6$ single crystals. Linearly dispersive bands were observed in its insulating gap around $\overline{\Gamma}$ and $\overline{X}$ with negligible $k_z$ dependences, indicative of their surface origins. The CD  of these in-gap states at various photon energies show patterns consistent with the locked spin-momentum texture of TIs. The CD pattern of the $\alpha$ band around $\overline{\Gamma}$ shows obvious two-fold symmetry, while the $\beta$ band around $\overline{X}$ presents anti-symmetric pattern about the $\overline{\Gamma}$-$\overline{X}$ axis in the surface Brillouin zone (BZ).  More importantly, we found that the chemical potential varies up to about 500~meV, from one cleaved surface to another, or sometimes even across the same surface. The bulk bands often coexist with the topological surface states, however, we also identified a bulk insulating gap of about 100~meV, much larger than predicted \cite{Ybcalc}. It is thus possible to tune the chemical potential in such a gap and enable applications based on the surface states.

High quality YbB$_6$ single crystals were synthesized by the Al-flux method as elaborated in the Supplementary Material (SM), together with their characterizations.  ARPES measurements were performed at the SIS beamline of Swiss Light Source (SLS) and Beamline 5-4 of Stanford Synchrotron Radiation Lightsource (SSRL), both equipped with a Scienta R4000 electron analyzer, and beamline 28A of Photon Factory, KEK, equipped with a Scienta SES-2002 electron analyzer. The angular resolution was 0.3$^\circ$ and the overall energy resolution was better than 15~meV depending on the photon energy. The samples were cleaved $in$ $situ$ along the (001) plane and measured under ultra-high vacuum below 5$\times$10$^{-11}$ torr.

Figure~\ref{cut}(a) shows the photoemission intensity of YbB$_6$ on sample 1 (S1) over a large energy scale, the non-dispersive Yb 4$f$ bands are located at around 1~eV and 2.3~eV below Fermi energy ($E_F$) . They correspond to the $^6$H$_{5/2}$ and $^6$H$_{7/2}$ multiplets of the Yb$^{2+}$ 4$f$$^6$-4$f$$^5$ final states based on the LDA calculation \cite{Ybcalc}. The energy difference between the two multiplets is 1.3~eV, consistent with the calculation. However, the energy positions of these 4$f$ bands are deeper than that calculated in Ref.~\onlinecite{Ybcalc}, probably due to the chemical potential shift induced by carrier doping. In addition, the dispersive feature centered at $\overline{\Gamma}$ between the two 4$f$ bands should originate from the Yb 5$d$ band.

The photoemission intensity map at $E_F$ on S1 is shown in Fig.~\ref{cut}(b). An elliptical Fermi pocket located around $\overline{X}$ and a circular Fermi pocket around $\overline{\Gamma}$ are clearly observed.  The photoemission intensity distribution across  $\overline{\Gamma}$  taken on sample 2 (S2) is plotted in Fig.~\ref{cut}(c), together with a background-substracted one to visualize the dispersions more clearly. There is an electron-like band  (referred to as $\alpha$),  forming the circular Fermi pocket in Fig.~\ref{cut}(b). In Fig.~\ref{cut}(e),  another  electron-like band is observed around $\overline{X}$ (referred to as $\beta$), forming  the elliptical Fermi pocket.

Both the $\alpha$ and  $\beta$ bands exhibit almost linear dispersions as can be seen from their momentum distribution curves (MDCs) in Figs.~\ref{cut}(d), 1(f) and 1(g). The $\alpha$  band extrapolates to cross at about 380~meV below $E_F$. There is also a $\gamma$ band below the $\alpha$ band. However, the band bottom of $\alpha$ overlaps with the band top of $\gamma$ in  certain energy range as highlighted in Fig.~\ref{cut}(c). This observation, together with the very different intensities of $\gamma$ and $\alpha$,  indicates that $\gamma$ is not the lower cone of  $\alpha$, if there is a Dirac point as predicted \cite{Ybcalc}.  Due to the low intensity of the $\alpha$ band and the interference of the strong $\gamma$ band, the lower cone  is hardly resolved, if any.

The $\beta$ band also seems to cross at about 320~meV below $E_F$ by tracking its linear dispersion, but the predicted lower cone could not be distinguished either, due to the strong residual intensity of the flat 4$f$ bands. The $\delta$ band is observed with a different photon energy on sample 3 (S3) with its band bottom around -80~meV [Fig.~\ref{cut}(h)]. The band bottom of $\beta$ in S3 is around -250~meV, different from that in S2, which may be caused by different chemical potentials. Figure~\ref{cut}(j) is the schematic of all the low-lying energy bands resolved along $\overline{\Gamma}$-$\overline{X}$ direction.

\begin{figure*}[t]
\includegraphics[width=18cm]{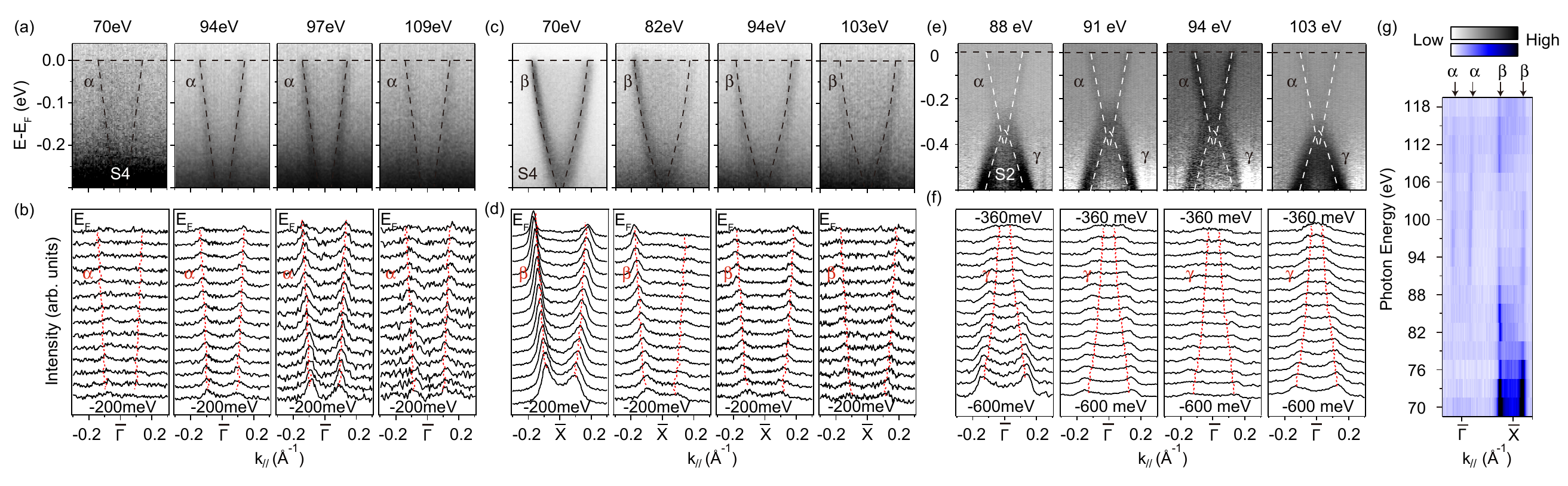}
\caption{(Color online) (a) Photon energy dependence of the photoemission intensity of the $\alpha$ band measured along cut 2 on S4. (b) The corresponding MDC plots of $\alpha$ in panel (a) within [$E_F$-200~meV, $E_F$]. (c), (d) The same for the $\beta$ band as in panel (a) and (b) measured along cut 3 on S4. (e) Photon energy dependence of the photoemission intensity of the $\gamma$ band measured along cut 2 on S2. (f) The corresponding MDC plots of the $\gamma$ band in panel (e) within [$E_F$-600~meV, $E_F$-360~meV]. The red dashed lines in panel (b), (d) and (f) are fitted dispersions with 94~eV data of the $\alpha$ and $\beta$ bands and with 103~eV data of the $\gamma$ band, respectively, and overlaid them onto the data of other photon energies. The white dashed lines in panel (e) are extracted from the 103eV data indicating the $\alpha$ and $\gamma$ bands, and overlaid them on top of data with other photon energies. (g) The $h\nu$-$k_\varparallel$ intensity plot of the MDCs at $E_F$ of various photon energies. The intensity was integrated in a window of [$E_F$-10~meV, $E_F$+10~meV]. Data were taken at 18 K at SLS.
} \label{kz}
\end{figure*}

Photon energy dependent ARPES experiments have been performed to examine how the $\alpha$, $\beta$ and $\gamma$ bands vary with $k_z$. The photoemission intensity distributions taken with different photon energies are presented in Fig.~\ref{kz}. One can see that, for both  $\alpha$ and $\beta$ bands, although the spectral weight varies with photon energy due to photoemission matrix element effects, the dispersions of them remain unchanged. Figures~\ref{kz} (a) and 2(b) show the photoemission intensity distributions across $\overline{\Gamma}$ on sample 4 (S4) together with their corresponding MDCs. We fitted the dispersion of the $\alpha$ band taken with 94~eV photons and overlaid it onto the data taken with several other photon energies, which clearly show that the dispersions of the $\alpha$ band are $k_z$ independent. The same process has been done to the $\beta$ band in Figs.~\ref{kz}(c) and 2(d), which also show no $k_z$ dependence of the $\beta$ band. More data under other photon energies can be found in the SM. Furthermore, the $h \nu$-$k_\varparallel$ intensity plot  at $E_F$ shown in Fig.~\ref{kz}(g) demonstrates the two-dimensional nature of the $\alpha$ and $\beta$ bands.

However,  in Figs.~\ref{kz}(e) and ~\ref{kz}(f), when we overlaid the  dispersion of  $\gamma$  extracted from the MDCs taken with 103~eV photons on top of data taken with several other photon energies, there are clear mismatches. That is,  the $\gamma$ band disperses with $k_z$, indicative of its   bulk origin. Furthermore, the dark interior of intensity envelopes of the $\gamma$ band in Fig.~\ref{kz}(e)  is likely  due to its dispersion along $k_z$ and the poor $k_z$ resolution in ARPES measurements with vacuum ultraviolet photons. Similar behavior has been reported for a bulk band  in SmB$_6$ \cite{JJSmB6}.

In addition, the $\delta$ band is only observed in certain photon energies, due to different cross-section of this band. The lineshape of the $\delta$ band is hardly resolved due to its bulk origin and poor $k_z$ resolution, which result in a bright interior of intensity envelopes of the $\delta$ band. Thus the $\delta$ band should be the bulk conduction band. As the band bottom of the $\delta$ band in S3 is around -80~meV, the band bottom of $\delta$ in S2 should be around -150~meV, due to different chemical potential between these two samples (see SM). Moreover, since the $\gamma$ band top in the 91 eV data of S2 locates at around -250~meV [Fig.~\ref{cut}(c)], the bulk band gap of YbB$_6$ is likely to be about 100~meV, as illustrated in Fig.~\ref{cut}(j). This determined gap size is much larger than the theoretical prediction of only 31~meV \cite{Ybcalc}, which suggests the correlation effect   needs to be better treated.

The chirality of spin and orbital angular momentum (OAM) is another remarkable hallmark of the topological surface state, since both of them are interlocked and rotate with the electron momentum, so that the total angular momentum is conserved.  To further explore the topological nature of these surface states, we conducted CD-ARPES experiment, which probes the chirality of the OAM.\cite{Changyoung,JJSmB6}

\begin{figure}
\includegraphics[width=8.7cm]{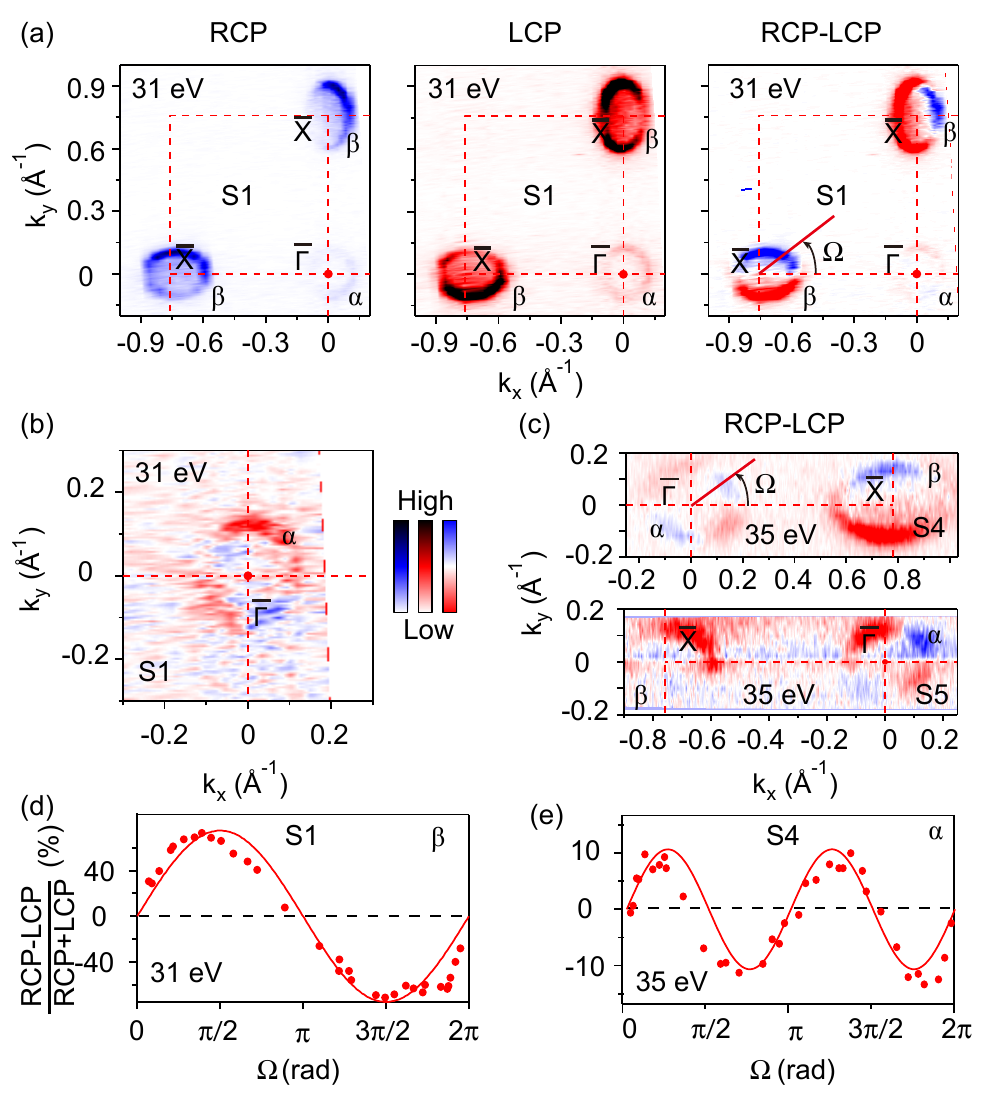}
\caption{(Color online) (a) Fermi surface maps of YbB$_6$ taken with RCP, LCP light and their difference (RCP-LCP) of S1. The intensity was integrated over a window of [$E_F$ - 10~meV, $E_F$ + 10 ~meV].  (b) The RCP-LCP data around $\overline{\Gamma}$ which is enlarged from panel (a). (c) Top: CD result of S4 (top) and S5 (bottom) measured with 35~eV photons showing the right $\overline{X}$ pocket and the left $\overline{X}$ pocket, respectively. (d) The CD values taken along the left pocket around $\overline{X}$ in panel (a) of S1, where the polar angle $\Omega$ is defined within panel (a) respect to $\overline{X}$. (e) The CD values of the $\alpha$ pocket taken around $\overline{\Gamma}$ of S4, where the polar angle $\Omega$ is defined within panel (c) respect to $\overline{\Gamma}$. The red solid curves in panel (d) and (e) are sine function fits of the data. The 31 ~eV data were taken at 6~K in SSRL and the 35 ~eV data were taken at 18~K in SLS.
 }
\label{CD}
\end{figure}

The Fermi surface maps of S1 measured under right-handed circular polarized (RCP) light and left-handed circular polarized (LCP) light are shown in Fig.~\ref{CD}(a), together with their difference (referred to as RCP-LCP). The CD of the $\beta$ pocket shows an anti-symmetric pattern about the $\overline{\Gamma}$-$\overline{X}$ axis in the surface BZ. The enlarged  CD image near $\overline{\Gamma}$ is shown in Fig.~\ref{CD}(b). Although its intensity is much weaker than that of the $\beta$ pocket, we can see that the symmetry of the $\alpha$ pocket becomes two-fold, which is different from the anti-symmetric CD pattern of the $\beta$ pocket. Such symmetries are confirmed by another set of data with more statistics  in the top panel of Fig.~\ref{CD}(c), which was measured  on S4  at a different photon energy. The bottom panel in Fig.~\ref{CD}(c) is measured on sample 5 (S5), under the same condition as S2, but presents the left $\beta$ pocket.  Clearly, all the CD patterns exhibit an inversion symmetry with respect to $\overline{\Gamma}$. Moreover, there is a sign change between the 31~eV and 35~eV CD data in both the $\beta$ and $\alpha$ pockets. The simultaneous sign change can be attributed to a final-state effect of the CD-ARPES  as reported in Bi$_2$Se$_3$ \cite{BiTeCD}. In Figs.~\ref{CD}(d) and 3(e), we plot the representative CD distributions around the $\alpha$ and $\beta$  pockets. Both of them can be well fitted by a sine function, however, the period for  $\beta$  is 2$\pi$, and that for  $\alpha$ is $\pi$. The reductive period of the $\alpha$ band, or the two-fold symmetric CD pattern needs further investigation, but similar behaviors were  attributed to  the coupling to  bulk states in Bi$_2$Se$_3$ and Bi$_2$Te$_3$ \cite{Jung, Bahramy, CD_theory, YHWang_2}. Despite the complexity, the CD results do suggest that the chirality of OAM exist in both the $\alpha$ and $\beta$ bands, indicating that they might possess helical spin textures, further support their topological origin.

\begin{figure}
\includegraphics[width=8.6cm]{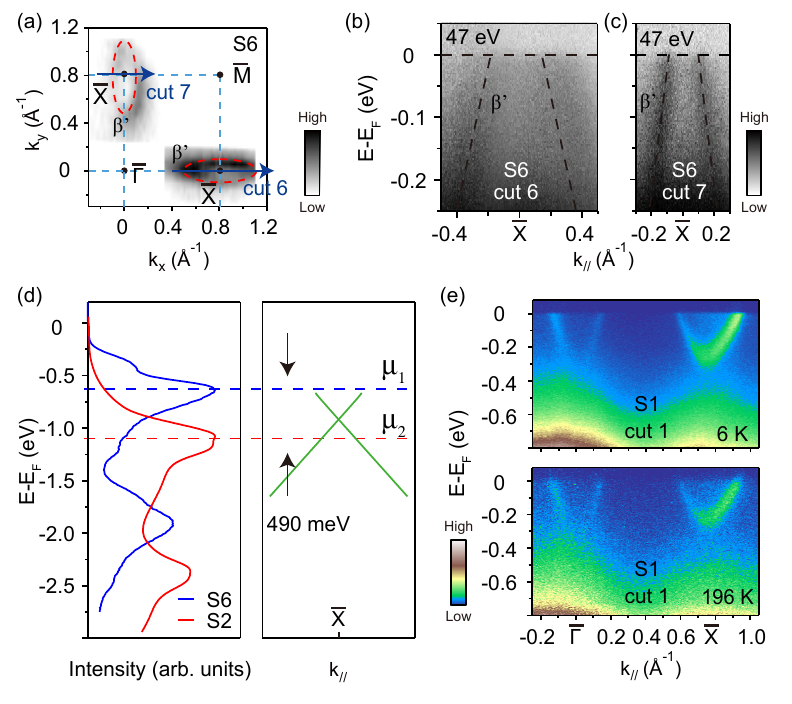}
\caption{(a) The photoemission intensity map of S6 showing hole-like pockets. The intensity was integrated over a window of [$E_F$ - 10~meV, $E_F$ + 10 ~meV]. (b), (c) Two photoemission intensity plots taken along cut 6 and cut 7 which is indicated in panel (a) by the blue arrows, respectively. The hole-like dispersions are indicated by the dashed lines. Data were taken on S6 at 20~K with 47~eV photons in KEK. (d) A schematic plot to show the chemical potential shift in different surfaces and the integrated EDCs of the two different surface to compare the different chemical potential. $\mu_1$ and $\mu_2$ refer to the chemical potential of the two different cleaved surfaces of S2 and S6. (f) The photoemission intensity data taken along $\overline{\Gamma}$-$\overline{X}$ direction at 6~K and 196~K, respectively. Data were taken on S1 at SSRL with 31~eV photons. } \label{hole}
\end{figure}

Having established the surface nature and the possible topological origin of the $\alpha$ and $\beta$ bands, we briefly discuss the Z$_2$ topology that can be determined from these bands. Since we observed one Fermi pocket around $\overline{\Gamma}$ and two Fermi pockets around $\overline{X}$ in the surface BZ, the total number of surface states is odd, thus the whole system should be topologically non-trivial. This is remarkably consistent with the theoretical prediction that YbB$_6$ is a TI \cite{Ybcalc}.


For a  surface state, the work function depends directly on the surface conditions. In various cleavages, we observed that the work function of YbB$_6$ varies about 100~meV, exposing the electronic structures discussed above. However,  we observed a  hole-like pocket around $\overline{X}$ twice,  referred to as $\beta'$  in Figs.~\ref{hole}(a)-4(c). Sometimes, the electron-like Fermi pocket and the hole-like Fermi pocket can even be observed in different regions of the same sample, which requires  further elucidation by scanning tunneling microscopy. The linear extrapolation of the $\beta'$ band leads to a crossing at around 180~meV above $E_F$. The slope of $\beta$ near its crossing point, and the observed slope of  $\beta'$ only differs about 4~\%, therefore $\beta'$ should be the lower half of the Dirac cone of   $\beta$. Since the Dirac point of  $\beta$ is located at around 320~meV below $E_F$, the shift of the chemical potential is about 500~meV, which equally affects the 4$f$ band as shown in Fig.~\ref{hole}(d). On the other hand, due to the weak intensity of the $\alpha$ band, we did not resolve its lower cone.

The observed surface states in YbB$_6$ is robust  against the change of temperature from 6~K to 196~K  [Fig.~\ref{hole}(e)]. This is consistent with the DFT calculations which show that YbB$_6$ is not a topological Kondo insulator \cite{Ybcalc}.

To summarize, we have obtained a comprehensive understanding of the electronic structure of YbB$_6$ from ARPES. We directly observed several linearly dispersive bands within the insulating gap of YbB$_6$ which show negligible $k_z$ dependence. The CD-ARPES data suggest well-defined  helical spin texture of the surface states. These results suggest that YbB$_6$ is a TI. In addition, the insulating gap observed here is much larger than that reported in theoretical calculations, indicating correlation effects. We note that the sensitivity of work function reported here also implies its tunability. If the work function can be tuned into the 100~meV insulating gap, the topological surface state can be exploited in applications without the interference from the bulk states.

We thank Dr. Ming Shi and Dr. N. Xu at SLS and Dr. K. Ono and Dr. N. Inami at KEK. This work is supported in part by the National Science Foundation of China and National Basic Research Program of China (973 Program) under the grant Nos. 2012CB921402, 2011CB921802, 2011CBA00112. SSRL is operated by the U. S. DOE Office of Basic Energy Science.

\end{document}


\title{Supplementary Material: Angle-resolved Photoemission Spectroscopy Study on the Surface States of  the Correlated Topological Insulator YbB$_6$}

\author{M. Xia \footnote{These authors contribute equally to this work.}}

\author{J. Jiang \footnote{These authors contribute equally to this work.}}

 \author{Z. R. Ye }

\affiliation{State Key Laboratory of Surface Physics, Department of Physics, and Advanced Materials Laboratory, Fudan University, Shanghai 200433, China}

\author{Y. H. Wang}

\affiliation{Department of Physics and Applied Physics, Stanford University, Stanford, California, 94305, USA}

\author{Y. Zhang}

\affiliation{SLAC National Accelerator Laboratory, Stanford Institute for Materials and Energy Sciences,
2575 Sand Hill Road, Menlo Park, California 94025, USA}

\author{S. D. Chen}

\affiliation{Department of Physics and Applied Physics, Stanford University, Stanford, California, 94305, USA}

 \author{X. H. Niu }

\author{D. F. Xu  }

\affiliation{State Key Laboratory of Surface Physics, Department of Physics, and Advanced Materials Laboratory, Fudan University, Shanghai 200433, China}

\author{F. Chen}
\author{X. H. Chen}
\affiliation{Department of Physics and Hefei National Laboratory for Physical Science at Microscale, University of Science and Technology of China, Hefei 230026, China.}

\author{B. P. Xie }

\author{T. Zhang }

\email{tzhang18@fudan.edu.cn}

\author{D. L. Feng }

\email{dlfeng@fudan.edu.cn}

\affiliation{State Key Laboratory of Surface Physics, Department of Physics, and Advanced Materials Laboratory, Fudan University, Shanghai 200433, China}


\maketitle

\section{Sample Characterization}

\begin{figure}
\includegraphics[width=8.7cm]{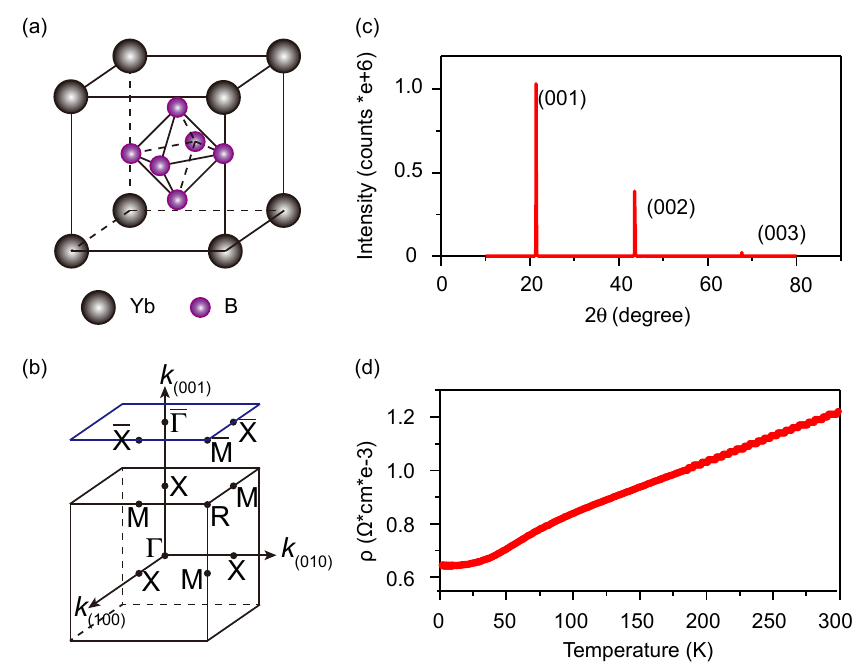}
\caption{(Color online) (a) Crystal structure of YbB$_6$.  (b) Bulk and surface Brillouin zone of YbB$_6$ and the high symmetry points. (c) X-ray diffraction pattern of YbB$_6$ single crystals. (d) Resistivity of YbB$_6$ single crystals measured from 2~K to 300~K.
 }
\label{sample}
\end{figure}

High quality YbB$_6$ single crystals were synthesized by the Al-flux method. A chunk of Yb together with the powders of Boron and Al were mixed with a ratio of 1:6:400 and heated to 1773~K in an alumina crucible in the argon atmosphere, then maintained at 1773~K for about 2 days before slowly cooling down to 873~K at a rate of 5~K per hour.

\begin{figure}
\includegraphics[width=8.7cm]{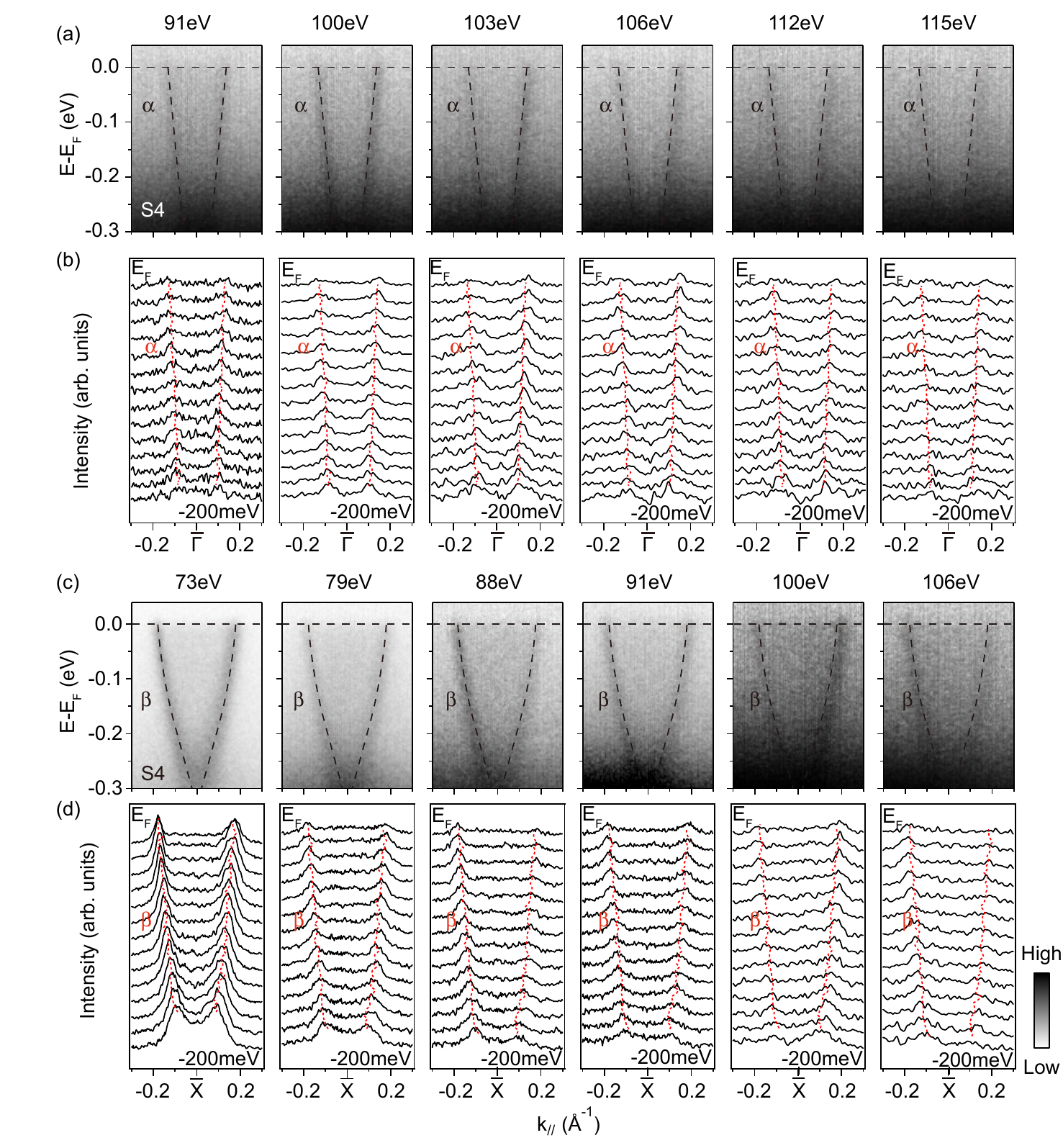}
\caption{(Color online) (a) Photon energy dependence of the photoemission intensity measured along cut 2 around $\overline{\Gamma}$ on S4, and (b), their corresponding momentum distribution curves (MDCs) within [$E_F$-200~meV, $E_F$]. The black dashed lines in panel (a) indicate the $\alpha$ bands and the red dashed lines in panel (b) are dispersions of the $\alpha$ band fitted from the 94~eV data, and overlaid them on data taken with other photon energies. (c), (d) The same as those in panel (a) and (b) to show the photon energy dependence of the $\beta$ band, measured along cut 3 around $\overline{X}$. Data were taken at 18~K at SLS.
 }
\label{kz}
\end{figure}

YbB$_6$ crystallizes in the cubic \emph{Pm-3m} space group and can be considered as the CsCl-type structure with the Yb ions and the B$_6$ octahedra being located at the corner and  the body center of the cubic lattice, respectively, as shown in Fig.~\ref{sample}(a). Figure ~\ref{sample}(b) shows the bulk Brillouin zone (BZ) together with the projected surface Brillouin zone. The high symmetry points we used are indicated in this panel. The X-ray diffraction pattern presented in Fig.~\ref{sample}(c) indicates a high sample quality of our single crystal and the natural surface is the (001) surface. The resistivity of YbB$_6$ single crystal shows a metallic behavior down to 2~K, which can be understood by the coexistence of the bulk band with the surface states.

\section{$k_z$ dependence of the $\alpha$ and $\beta$ bands}

To prove the surface origin of the $\alpha$ and $\beta$ bands, photon energy dependent measurements were performed to study their evolution with $k_z$. Photon energies in a large energy scale (70~eV-118~eV) were used to cover the BZ along the $k_z$ direction, as $k_z$ is related to photon energy ($h\nu$) through $k_z$=0.512$\sqrt{h\nu-W+V_0}$, where $W$ is the work function and $V_0$ is the inner potential. If we take a typical value of $V_0$ as 15~eV, the total photon energies cover 0.8~BZ along the $k_z$ direction in the three-dimensional BZ, which is enough to see the periodicity of the $\alpha$ and $\beta$ bands from $\Gamma$ to $Z$.

In addition to the data of the $\alpha$ and $\beta$ bands taken under several photon energies shown in the main text, data under more photon energies can be found in Fig.~\ref{kz}. All our data strongly support the surface origin of both the $\alpha$ and $\beta$ bands.


\begin{figure}[t]
\includegraphics[width=8.7cm]{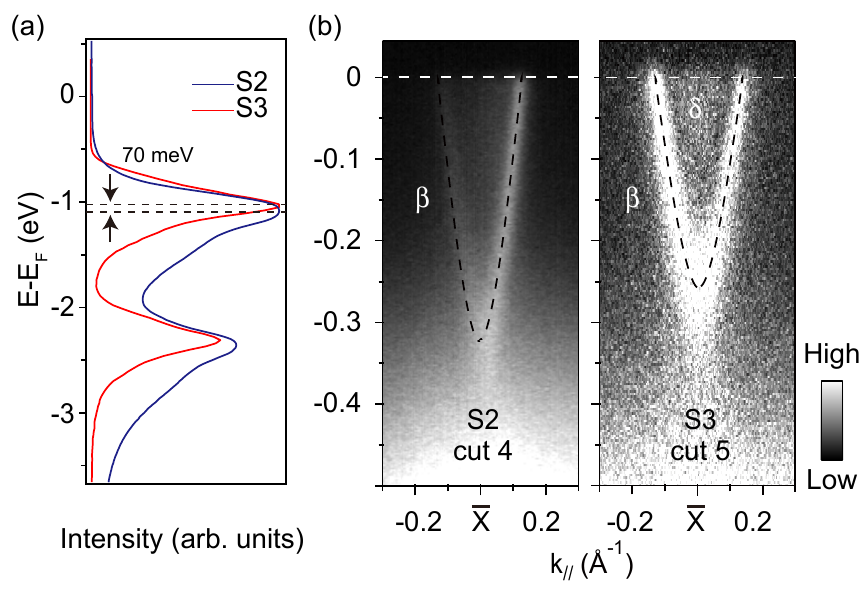}
\caption{(Color online) (a) The integrated large scale energy distribution curves (EDCs) of the two different samples to compare the different chemical potential. (b) The photoemission intensity plot on S2 along cut 4 and S3 along cut 5.  The data of S2 were taken at 18~K with 70~eV photons in SLS, and the data of S3 were taken at 20~K with 47~eV photons in KEK.
 }
 \label{VB}
 \end{figure}

\section{Chemical potential difference between different samples}

The work function of TI depends directly on the surface conditions. For our most samples, the work function varies less than 100~meV. From the valence band spectra of S2 and S3 shown in Fig.~\ref{VB}(a), we can see that the 4$f$ band shifted around 70~meV. This is consistent with that we determined from the band bottom of the $\beta$ band in these two samples in the main text. One would notice that the Fermi crossing of the $\beta$ band in S3 is a little larger than that in S2, this is because that cut 5 is 3.5$^\circ$ off cut 4, hence it is not exact along the minor axis of the elliptical $\beta$ pocket.